\documentclass[aip,amsmath,amssymb,preprint]{revtex4-2}
\usepackage[pass]{geometry}
\usepackage{graphicx}
\graphicspath{ArXiv_Submission_04Feb2022/ArXiv_04Feb2022}
\usepackage{dcolumn}
\usepackage{bm}
\usepackage{textcomp}
\usepackage{verbatim}
\usepackage{siunitx}
\usepackage[utf8]{inputenc}
\usepackage[T1]{fontenc}
\usepackage{mathptmx}
\usepackage{gensymb}
\usepackage[compact]{titlesec}
\titlespacing{\section}{0pt}{1ex}{0ex}
\titlespacing{\subsection}{0pt}{1ex}{0ex}
\newcommand{\beginsupplement}{
        \setcounter{table}{0}
        \renewcommand{\thetable}{S\arabic{table}}
        \setcounter{figure}{0}
        \renewcommand{\thefigure}{S\arabic{figure}}
        \setcounter{section}{0}
        \renewcommand{\thesection}{S\arabic{section}}
        \setcounter{equation}{0}
        \renewcommand{\theequation}{S\arabic{equation}}
     }
    
\begin{document}
\title{\Large\begin{center} Three-axis torque investigation of interfacial exchange coupling in a NiFe/CoO bilayer micromagnetic disk \end{center}}

\author{M.G. Dunsmore}
\thanks{M.G. Dunsmore and J.A. Thibault contributed equally to this work.}
\affiliation{University of Alberta, Department of Physics, Edmonton, Alberta, T6G 2E1, Canada}
\author{J.A. Thibault}
\thanks{M.G. Dunsmore and J.A. Thibault contributed equally to this work.}
\affiliation{University of Alberta, Department of Physics, Edmonton, Alberta, T6G 2E1, Canada}
\author{K.R. Fast}
\affiliation{University of Alberta, Department of Physics, Edmonton, Alberta, T6G 2E1, Canada}
\author{V.T.K Sauer}
\affiliation{University of Alberta, Department of Physics, Edmonton, Alberta, T6G 2E1, Canada}
\author{J.E. Losby}
\affiliation{University of Alberta, Department of Physics, Edmonton, Alberta, T6G 2E1, Canada}
\affiliation{University of Calgary, Department of Physics and Astronomy, Calgary, Alberta, T2N 1N4, Canada}
\affiliation{Nanotechnology Research Centre (NANO), National Research Council Canada (NRC), Edmonton, Alberta, T6G 2M9, Canada }
\author{Z.Diao}
\affiliation{University of Alberta, Department of Physics, Edmonton, Alberta, T6G 2E1, Canada}
\affiliation{Maynooth University, Department of Electronic Engineering, W23 F2H6 Maynooth, County Kildare, Ireland}
\author{M. Belov}
\affiliation{Nanotechnology Research Centre (NANO), National Research Council Canada (NRC), Edmonton, Alberta, T6G 2M9, Canada }
\author{M.R. Freeman}
 \email{freemanm@ualberta.ca}
\affiliation{University of Alberta, Department of Physics, Edmonton, Alberta, T6G 2E1, Canada}

\begin{abstract}
    Micrometer diameter bilayers of NiFe (permalloy, Py) and cobalt oxide (CoO) deposited on nanomechanical resonators were used to investigate exchange bias effects. The mechanical compliances of two resonator axes were enhanced by severing one torsion arm, resulting in a unique three-axis resonator that responds resonantly to torques generated by a three-axis RF field. Our technique permits simultaneous measurement of three orthogonal torque components. Measurements of the anisotropies associated with interfacial exchange coupling effects have been made. At cryogenic temperatures, observations of shifted linear hysteresis loops confirmed the presence of exchange bias from the Py/CoO interface. An in-plane rotating DC bias field was used to probe in-plane anisotropies through the out-of-plane torque. Training effects in the rotational hysteresis data were observed and showed that features due to interfacial coupling did not diminish irrespective of substantial training of the unidirectional anisotropy. The data from the rotational hysteresis loops were fit with parameters from a macrospin solution to the Landau-Lifshitz-Gilbert equation. Each parameter of the exchange bias model accounts for specific features of the rotational loop. 
\end{abstract}

\date{04 February 2022}
\maketitle

\section{Introduction}
Magnetic torque is a probe of magnetic anisotropy. We examined a thin film bilayer of ferromagnetic (FM) Py and antiferromagnetic (AF) CoO with a three-axis AC torque magnetometer\cite{Fast2021}. The multi-axis measurements enabled extraction of all pertinent magnetic information without requiring additional samples or a separate magnetometer. Multi-axis mechanical resonators have been useful in previous studies of Py at room temperature\cite{Mattiat2020,Fast2021}. A notable capability of the three-axis technique is that the saturation magnetization, $M_{s}$, can be determined during a single measurement of the in-plane magnetization. Thin-film magnetic structures have strong out-of-plane shape anisotropy and, with in-plane magnetization, have in-plane torques that provide a measure of the object's magnetic moment. The out-of-plane torque is dominated by interfacial exchange coupling between the Py and CoO.

Exchange bias refers to effects originally observed by Meiklejohn and Bean in 1956 \cite{meiklejohn1957}. Since then, there has been considerable interest in the topic ranging from pure scientific endeavor to specific application, (e.g. spin valves in hard drives). There are many publications\cite{nolting2000,jimenez2009,maniv2021} and review articles\cite{kiwi2001,nogues2005} spanning the time frame from original discovery to more recent work\cite{dias2014,mitrofanov2021}. We found that rotating hysteresis was more informative than linear, a result seemingly in accordance with Meiklejohn's 1962 comments\cite{meiklejohn1962} regarding rotational hysteresis loss as more fundamental than shifts in linear hysteresis.

Experimental data were analyzed and compared to simulations derived from macrospin solutions to the Landau-Lifshitz-Gilbert equation. We observed distinct effects in rotational hysteresis loops attributable to interfacial exchange coupling that were independent of shifts in linear loops and were not explainable with unidirectional anisotropy as the dominant mechanism, consistent with other recent findings\cite{dias2014,mitrofanov2021}. Both high and zero field cooling caused an increase of in-plane anisotropy resulting from interfacial coupling that persisted after training had substantially reduced the exchange bias\cite{schlenker1986}.  

\section{Experimental Details}
\subsection{Sample}
Refer to Figure 1, panel (a). Doubly-clamped nanomechanical resonators were fabricated in a silicon-on-insulator wafer. To improve mechanical susceptibility to torques along the $x$- and $z$-axes, one torsion arm was cut with a focused ion beam. CoO was sputtered on the resonator paddle followed by Py (both layers 20 nm thick, Py not capped. See Supplementary Material Section S1). After lift-off, the as-patterned diameter of the bilayer disk was 1.36 $\si{\mu}$m as measured by scanning electron microscopy.

\subsection{Apparatus}
The sample wafer is mounted in a cryostat (4 to 300 K). A fixed RF coil generating field strengths at the sample, similar in $x$ and $z$, but negligible in $y$, is driven by a lock-in amplifier providing three simultaneous drive frequencies corresponding to the resonator's fundamental frequencies (1.89, 4.23, and 0.96 MHz for $\tau_x$, $\tau_y$, $\tau_z$ respectively). A DC magnet provides the bias field. The field magnitude and direction can be adjusted by translation along a rail and by rotation of the magnet. The configuration results in generation of the required DC and AC fields for simultaneous AC dither of torques along the $x$-, $y$-, and $z$-axes. A complete description of the DC and RF fields and how the fields combine to produce torque are given in Supplementary Material Section S2. \par

The torques can be expressed as a combination of moment and susceptibility terms. For example,
\begin{equation}
\tau_y = -\mu_0 m_x H_z + \mu_0 V \chi_z H_z H_x,
\end{equation}
\noindent where $m_x$ is the $x$-component of magnetic moment, $V$ is the ferromagnetic volume and $\chi_z$ is the field-dependent dimensionless magnetic susceptibility in the $z$-direction.  The expression generalizes to the other torque components. The AC torque magnetometer provides a probe for measuring small perturbations of the magnetization direction via curvature of the magnetic free energy similar to ferromagnetic resonance\cite{mcmichael1998}. Finally, the AC torque signals are sensed interferometrically via the Fabry-Perot cavity formed between the resonator paddle and the substrate base. Conversion from voltage to torque is done by thermomechanical calibration wherein the response of the resonator to thermal excitation is measured and used as the calibration standard\cite{losby2012} (see Supplementary Material Section S3).

\begin{figure}[htb!] 
\centering
\includegraphics[width=1\columnwidth]{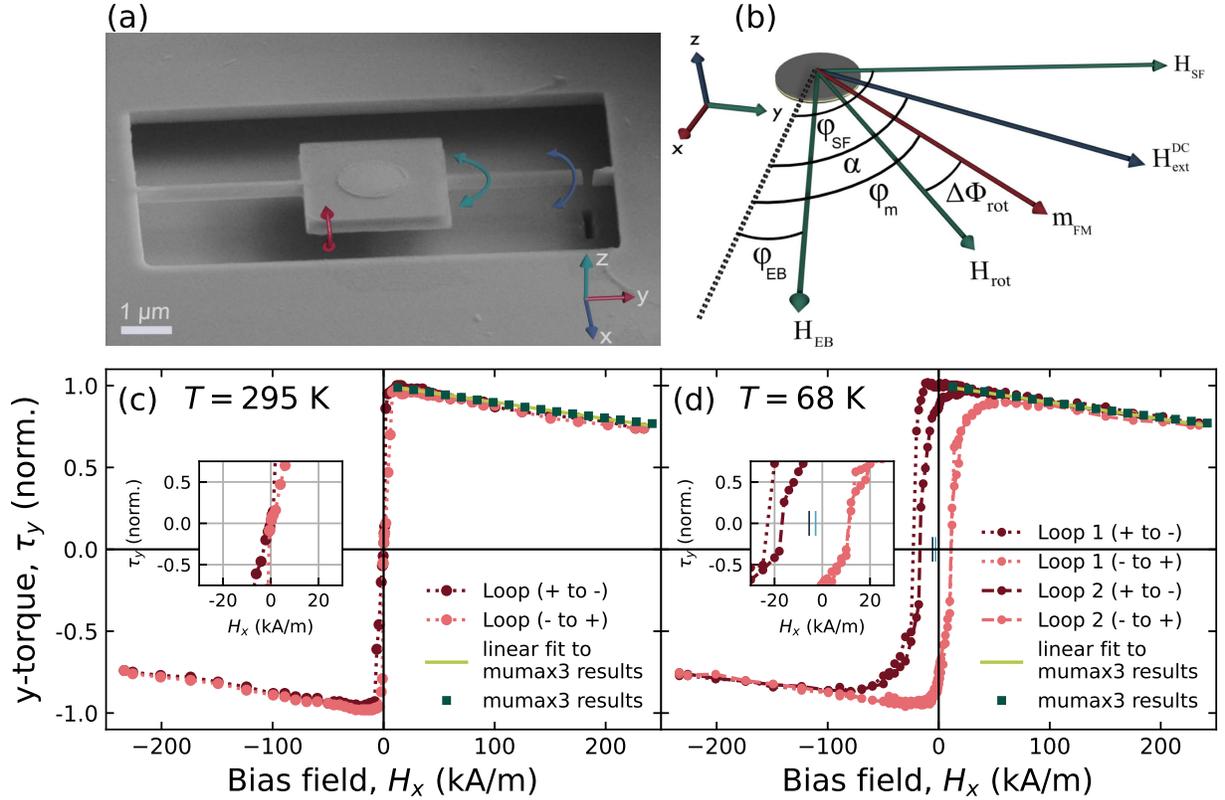}
\caption{Sample geometry, coordinate system, and sample characterization. (a), electron micrograph of resonator. Arrows indicate torque directions. (b), field vectors describing anisotropic torque contributions when cooled below the Néel temperature. $H_{\textrm{EB}}$, unidirectional exchange bias. $H_{\textrm{SF}}$, uniaxial spin-flop coupling. $H_{\textrm{rot}}$, rotatable anisotropy. $H_{\textrm{ext}}^{\textrm{DC}}$, external bias field. $m_{\textrm{FM}}$, Py magnetic moment. $H_{\textrm{EB}}$ is in the cooling field direction. Angles referenced to lab-frame $x$ except for $H_{\textrm{rot}}$ which is referenced to $m_{\textrm{FM}}$. (c) and (d), hysteresis loops above and below the Néel temperature. Insets show the coercive points. High field regions show a linear decrease in torque that was reproduced in Mumax3\cite{vansteenkiste2014}. A 16 nm Py magnetic thickness (see text) gave $M_{s}(295\textrm{ K})$ = 763 $\pm$ 7 kA/m and $M_{s}(68\textrm{ K})$ = 801 $\pm$ 8 kA/m (see Supplementary Material Section S5).} 
\label{fig:1comp}
\end{figure}

\section{Results and Discussion}
\subsection{Magnetic sample characterization and demonstration of exchange bias}
Figure 1 panel (c) shows a room temperature linear loop. The Néel temperature of CoO is roughly 290 K, so the data should not show a shift in coercive fields. The significant features in the plot are the narrow width near zero field, which is expected from Py, the absence of a shift in coercive points, and the negative slopes in the high field segments. These same data, when plotted against a reduced field range, show vortex nucleation and annihilation attributable to the known magnetic vortex behaviour of Py at room temperature\cite{burgess2013}.  The low-field magnetizing slope of the disk in the vortex state encodes the magnetic thickness of the permalloy, found to be $16 \pm 1\thinspace$nm through comparison with micromagnetic simulation of torque (see Supplementary Material Section S4).  That this is thinner than the 20 nm deposited thickness is consistent with oxidation of the uncapped surface.  

With the magnetization in saturation along $\hat{x}$, $\chi_z = M_s/(H_x + H_{\textrm{eff}})$, where the effective field from shape anisotropy is $H_{\textrm{eff}} = (N_z - N_r)M_s$ and $N_z, N_r, M_s$ are the demagnetizing factors (axial and radial) and saturation magnetization.  The negative slopes of the torque in saturation arise from the increase with field of the AC torque component due to $\chi_{z}$. An estimate of $M_s$ was extracted from the negative slopes (refer to Supplementary Material Section S5 for details). In the case at low temperature where anisotropy is induced by exchange coupling, the same procedure of micromagnetic simulation was followed with the addition of a uniaxial anisotropy. 

The insets in panels (c) and (d) show the zero crossings. The absence of a shift in panel (c) is an expected property of Py alone. The Py/CoO interface has not affected the room temperature behaviour of the Py.  Panel (d) shows first and second linear loops obtained after cooling to 68 K in a saturating DC bias field along the $x$ direction. Compared to the room temperature data in panel (c), the loops are wider and show a shift of the coercive fields. The decreased shift of the second loop is the result of training and is due mostly to a change in the zero crossing during the down sweep of the field. The up sweep zero crossings are much closer together. This feature has been observed by Qiu et al.\cite{qiu2008} and by Jenkins et al.\cite{jenkins2021}

\subsection{In-plane magnetization components and out-of-plane torque}

\begin{figure}[htb!] 
\centering
\includegraphics[width=0.5\columnwidth]{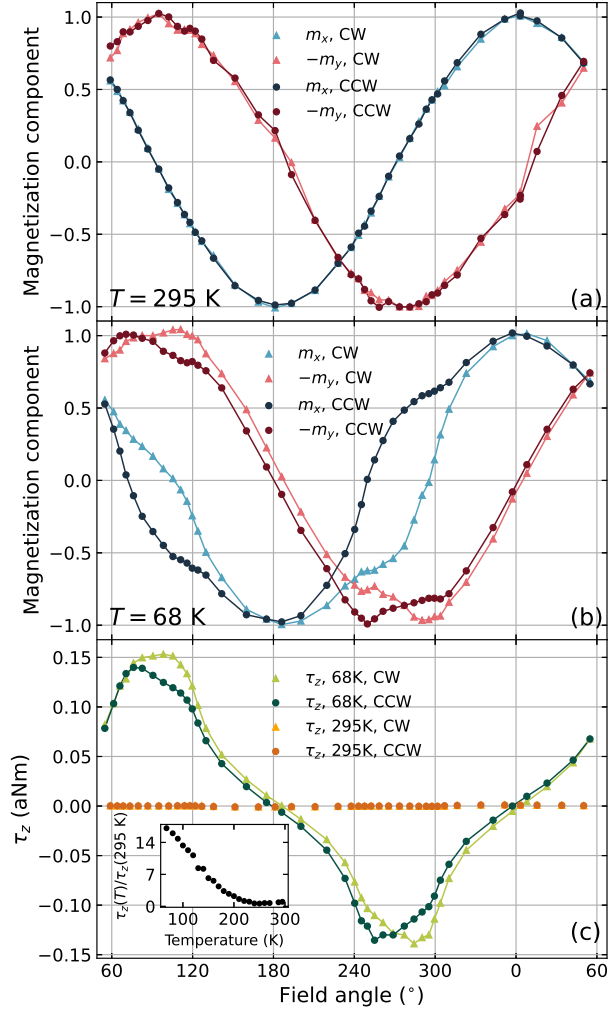}
\caption{Rotational loops reveal substantial differences in contributions to the total in-plane magnetic moment. (a), $m_{x}$ and $m_{y}$ for room temperature loops when exchange bias effects are not present. (b), magnetizations at 68 K show hysteretic behaviour not present in (a). (c), $z$-torque at 295 K and 68 K. The inset in panel (c) shows increase of torque with decreasing temperature, at a fixed field angle of $54^{\circ}$, due to the onset of in-plane anisotropy from exchange coupling. The inset $y$-axis is scaled to the room temperature value of $z$ torque (also at $54^{\circ}$ field angle).}
\label{fig:2}
\end{figure}

Figure \ref{fig:2} panel (a), room temperature, and panel (b), field cooled to 68$^{\circ}$ K, show the normalized in-plane magnetizations during rotational hysteresis loops that were obtained by rotation of the in-plane field angle through 360$^{\circ}$ counterclockwise (CCW) followed by rotation through 360$^{\circ}$ clockwise (CW). The maximum and minimum in-plane fields were 199.8 kA/m at 359.6$^{\circ}$ and 17.7 kA/m at 93$^{\circ}$ respectively (see Supplementary Material Section S2 and Fig. S1). $m_{x}$ and $m_{y}$ were obtained by correcting the $x$- and $y$-torques for the reduced output that occurred during the negative slope portion of the linear hysteresis curves in Figure 1 panels (c) and (d). The $y$-torque was used to calculate $m_{x}$. The $x$-torque was used to calculate $m_{y}$. Fitting at different temperatures yielded an accurate measure of the temperature dependence of the saturation magnetization.   

Figure 2 panel (a) shows that the $x$- and $y$-magnetizations repeated for the CCW and CW field rotations thus confirming that hysteresis due to anisotropies was not present in the room temperature data. The normalized magnetizations were combined to give the normalized saturation magnetization $m_{x}^{2} + m_{y}^{2} = 1$ ($m_{z}$ is negligible). This confirmed that the magnetization was in-plane and saturated during the entire 360$^{\circ}$ CCW and CW rotations. The field cooled data in panel (b) show that, while maintaining the correct relationship between the magnetizations, there was substantial hysteresis in the CCW and CW rotations attributable to an induced interaction between the Py and CoO brought about by field cooling.  

The room temperature data in Figure 2 panel (a) show that $m_{x}$ and $m_{y}$ rotated freely with the DC bias field. This occurred because the magnetocrystalline anisotropy in Py is very small and because the Py/CoO interface was inactive.  Panel (b) shows hysteresis in the field cooled in-plane magnetizations thus confirming the emergence of in-plane anisotropy. Correspondingly, Figure 2 panel (c) shows the $z$-torque during rotational loops at room temperature and after field-cooling. The torque is expressed in units of aNm via thermomechanical calibration\cite{losby2012} (details of the thermomechanical calibration are presented in Supplementary Materials Section S3). The field cooled $z$-torque shows an increase substantially above the room temperature values. The inset illustrates the nature of the temperature dependence of the torque with temperature, at a fixed in-plane field angle.

\subsection{Simulations}
\label{subsec:sim}
\begin{figure}[htb!] 
\centering
\includegraphics[width=0.5\columnwidth]{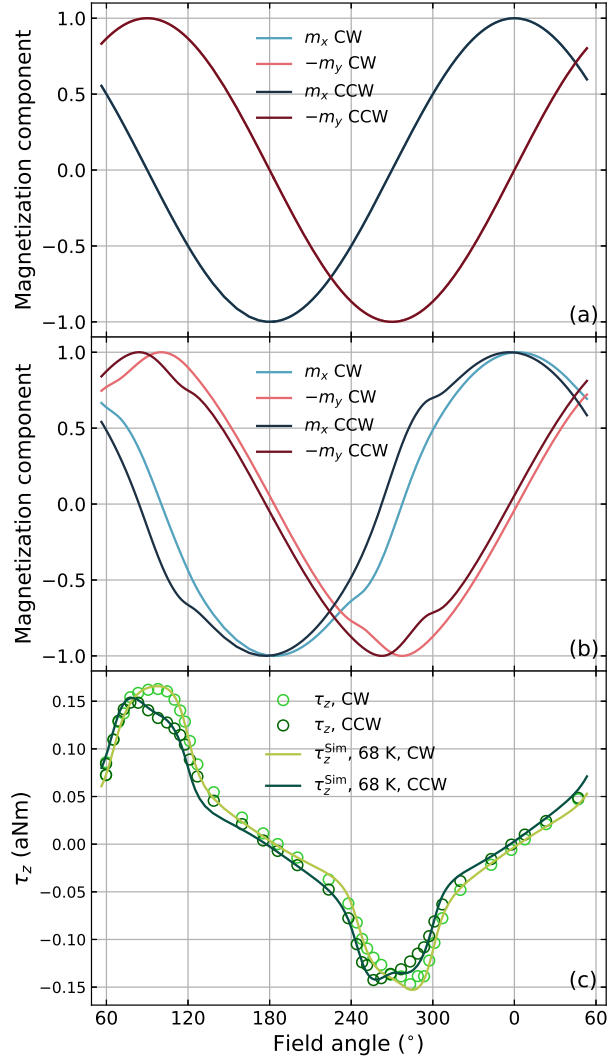}
\caption{Results from macrospin LLG solutions. Mirroring Figure \ref{fig:2}, (a) and (b) show the $x$- and $y$-magnetizations during field rotation. (a), no anisotropies aside from considerations of demagnetizing effects. (b), results after introduction of unidirectional, uniaxial, and rotatable anisotropies. (c), thermomechanically calibrated torque data with an overlaid fit from the macrospin model.} 
\label{fig:3}
\end{figure}

Quadrature addition of magnetization components from Figure 2 confirmed that the magnetization was in-plane and saturated for the entire field rotation. We can, therefore, neglect domain structure effects and model the influence of anisotropy on torque by solving the Landau-Lifshitz-Gilbert (LLG) equation for a single macrospin (macroscopic moment). The exchange coupling energy density is composed of unidirectional, spin-flop, and rotatable anisotropies, as in exchange bias models previously reported\cite{stiles1999,muglich2016,dasilva2018}. The unidirectional anisotropy, normally used to account for shifted linear loops, is created by the presence of uncompensated spins at the FM/AF interface, that is, spins that cannot rotate with the FM magnetization. Uncompensated spins are thought to maintain their positions along the field cooling axis and contribute to a net positive moment at the interface. The compensated spins that are still exchange coupled rotate along with the FM magnetization, giving rise to a rotatable anisotropy axis. The spin-flop coupling was introduced as a uniaxial anisotropy term. If we consider canting of AF sublattices away from their equilibrium positions, when the FM magnetization is 90$^{\circ}$ from the AF ordering axis, an easy axis forms since this is an energetically favourable alignment of coupled spins. We write the anisotropy energy density as follows,
\begin{equation}
   \begin{aligned}
     \varepsilon_{\textrm{A}}=&-K_{\textrm{SF}}\sin^{2}(\theta)\cos^{2}(\phi-\phi_{\textrm{SF}})-K_{\textrm{EB}}\sin(\theta)\cos(\phi-\phi_{\textrm{EB}})-K_{\textrm{rot}}\hat{m}_{\textrm{FM}}\cdot\hat{H}_{\textrm{rot}}
    \end{aligned}
\end{equation} 
where $K_{\textrm{SF}}$, $K_{\textrm{EB}}$, and $K_{\textrm{rot}}$ are the anisotropy constants for the spin-flop coupling, unidirectional exchange bias, and rotatable anisotropy respectively. $\phi_{\textrm{SF}}$ and $\phi_{\textrm{EB}}$ define the easy axis and easy direction. In addition to anisotropy arising from exchange coupling, the demagnetizing energy density, $\varepsilon_{\textrm{D}}$, was included to model the shape anisotropy of a ferromagnetic cylinder with cylindrical demagnetization factors, $N_{r}$ and $N_{z}$, found using equations derived by Joseph\cite{joseph1966}. The final energy density term is the Zeeman energy density, $\varepsilon_{\textrm{Z}}$, that describes the influence of the external magnetic field on the macrospin. A full description of the LLG equation and details of the simulation are included in Supplementary Materials Section S6 A. Figure 3 shows macrospin solutions to the LLG equation. Panel (a) shows the normalized in-plane magnetizations as a function of field angle and in the absence of in-plane anisotropy.  The normalized magnetizations repeat for the CCW and CW rotations and show no hysteresis effects. The results are very similar to the room temperature data in Figure 2 panel (a) and show that the simulation is adequate for reproducing the experimental observations. For simulations of the measurements after cooling in a 36 kA/m DC field at an in-plane angle of $55^{\circ}$, non-zero anisotropy values are required: a unidirectional anisotropy of $K_{\textrm{EB}}=(0.10\pm 0.05)$ kJ/m$^{3}$ along the cooling axis with $\phi_{\textrm{EB}}=55^{\circ}\pm 3^{\circ}$, a rotatable anisotropy axis with $K_{\textrm{rot}}=(22 \pm 1)$ kJ/m$^{3}$ and $\Delta \Phi_{\textrm{rot}}=-5^{\circ}\pm3^{\circ}$, and a uniaxial spin-flop coupling term with $K_{\textrm{SF}}=(8.3\pm 0.3)$ kJ/m$^{3}$ and $\phi_{\textrm{SF}}=-91^{\circ}\pm 2^{\circ}$. A best fit to the data with $\chi^2_{\nu,\mathrm{min}}$=1.2 was obtained using a standard calculation of $\chi^2_{\nu}$. The one-standard deviation contour of $\chi^2_{\nu}$ was subsequently used to bracket the uncertainties in the fit parameters (details of the fitting procedure and uncertainty calculations can be found in Supplementary Materials Section S6 D). The simulation provides magnetizations influenced by the anisotropies in a way that is meant to model the effects of exchange bias and interfacial coupling.

Hysteresis in the CCW and CW rotations is the relevant feature in the experimental data in Figure 2 panel (b). The simulated data in Figure 3, panel (b), shows similar hysteresis.  Figure 3 panel (c) compares the observed $z$-torque from a field cooled rotational loop to modeled torque. The important features in panel (c) are hysteresis in the CCW and CW rotations, asymmetry in the extrema magnitudes, and one-fold rotational symmetry. The simulated data agree well with experimental observations. Inspection of the simulation algorithm revealed that unidirectional anisotropy causes the asymmetry in extrema magnitudes, the spin-flop term influences the angular positions of the peaks, and orientation of the rotatable anisotropy axis is such that a one-fold rotational symmetry is introduced (additional details of the LLG equation and the rotatable anisotropy axis are presented in Supplementary Material Section S6 B).

\subsection{Comparison of first and trained loops}
\begin{figure}[htb!] 
\centering
\includegraphics[width=0.5\columnwidth]{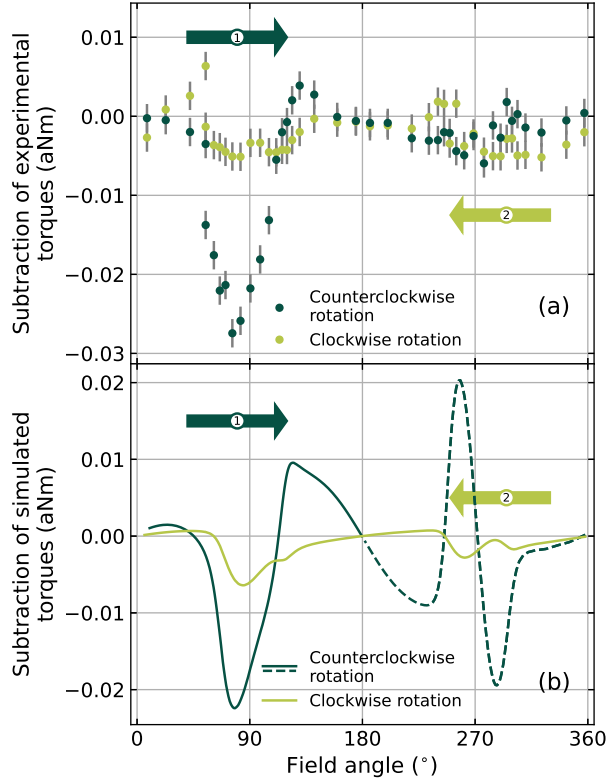}
\caption{Subtractions of first and highly trained loops of $z$-torque reveal subtleties otherwise obscured in the data. (a), difference between first loop following cooling and a significantly trained loop. Error bars are $\pm$ 0.002 aNm. The rotation begins at the cooling angle (0$^{\circ}$) and proceeds CCW followed by CW. A prominent fang-shaped spike appears in the beginning portion of the CCW rotation. (b), macrospin simulations. Solid curve shows region of CCW rotation where simulation correlates with experimental data. Dashed curve shows a second fang that does not appear in (a) because the simulation does not account for training effects.}
\label{fig:subtraction}
\end{figure}

Figure \ref{fig:subtraction} panel (a) shows the difference between $z$-torques from first and trained (eighth) rotational loops. The loops began with the CCW rotation, starting at 0$^{\circ}$ field angle, after cooling with the in-plane field strength set to 200 kA/m (at $0^{\circ}$). The difference between first and trained loops peaks near 75$^{\circ}$ after which the difference is less pronounced. This feature shows that substantial training occurred from 0$^{\circ}$ to 180$^{\circ}$. 

Figure 4 panel (b) shows simulation results for subtracted first and trained loops. Since the macrospin model has no mechanism for the training effect, the two measured loops were fit independently with their difference shown in panel (b). Significant training in the first half CCW rotation produced a large fang-shaped spike. Since the same parameters were used in the second half of the CCW rotation (dashed line), a second large spike occurs that does not appear in the data. The CW loop does not have any such effect suggesting that significant training occurred in the first half loop and that the result is due to athermal training, similar to observations by Qiu\cite{qiu2008}. The simulation parameters have a dramatic decrease in the unidirectional anisotropy from $K_{\textrm{EB}}=(1.00\pm0.05)$ kJ/m$^{3}$ in the CCW field rotation to $K_{\textrm{EB}}=(0.45\pm 0.05)$ kJ/m$^{3}$ in the CW rotation (both with $\phi_{\textrm{EB}}=0^{\circ}$).  The fractional change is similar to that found in the linear loop shifts in Fig. 1d, but, more significantly, the magnitudes are about $5\times$ smaller than what would be deduced by ascribing the entire loop shift to exchange bias alone.  Additionally, there is a slight increase in the spin flop anisotropy constant from $K_{\textrm{SF}}=(7.2\pm0.1)$ kJ/m$^{3}$ to $K_{\textrm{SF}}=(7.3\pm0.1)$ kJ/m$^{3}$. $K_{\textrm{rot}}$ remains the same at $K_{\textrm{rot}}= (22.5\pm0.3)$ kJ/m$^{3}$, but the orientation changes from $\Delta\Phi_{\textrm{rot}}=-3^{\circ}\pm 2^{\circ}$ to $\Delta\Phi_{\textrm{rot}}=3^{\circ}\pm 2^{\circ}$. The reduction in unidirectional anisotropy is responsible for the absence in the experimental data of the feature at 255$^{\circ}$ that appears in the model prediction. The dashed portion of the simulation is what would have appeared in the experimental data had there been no decrease in the unidirectional anisotropy.

The results indicate that continuous monitoring of training effects is obtainable from the out-of-plane torque response during rotating hysteresis loops whereas the change in shift of linear loops provides training information only after the first magnetization reversal thus obscuring earlier training effects. Finer angular step sizes than used in the current work ($9^{\circ}$) may produce data with more information regarding training.  

\section{Conclusions}
A bilayer film of Py/CoO, deposited on nanomechanical resonators, was used to study exchange bias and interfacial exchange coupling. A simultaneous three-axis AC torque magnetometer enabled investigations of interfacial exchange and exchange bias behaviours.

Examination of the in-plane magnetizations showed that a saturated magnetization state was present during all portions of rotational hysteresis loops thus allowing for a macrospin representation. A limitation of the macrospin model is its inability to address unsaturated states where micromagnetic techniques would be more appropriate.   

The rotational loops exposed more detail of interfacial coupling than the linear loops with effects independent of training that dominated over other contributions to in-plane anisotropy.

Further studies could include a more detailed description of the exchange coupled spin state. Additionally, the anisotropy temperature dependence could provide more information regarding AF grain size distributions\cite{muglich2016}.

\section*{Supplementary Material}
See Supplementary Material for additional details regarding sample fabrication, measurement apparatus, thermomechanical calibration, in-plane torque at low and high bias fields, experimental justification for the macrospin description, and LLG macrospin simulations.

\section*{Acknowledgments}
The authors gratefully acknowledge support from the Natural Sciences and Engineering Research Council of Canada (RGPINs 04239 and 2021-02762), the Canada Foundation for Innovation (34028), the Canada Research Chairs (230377).  The nanomechanical torque devices were created using tools in the University of Alberta nanoFAB and at the National Research Council Nanotechnology Research Centre.  We are indebted to Dr. Erik Luber for expert deposition of the bilayers.

\section*{Author Declarations}
The authors have no conflicts to disclose.

\section*{Data Availability}
Data are available from the corresponding author upon reasonable request.

\bibliography{Bibliography_Manuscript_19_Jan_2022.bib}
\newpage

\begin{center}
\Large\textbf {Three-axis torque investigation of interfacial exchange coupling in a NiFe/CoO bilayer micromagnetic disk: Supplementary Material}
\end{center}
M.G. Dunsmore, J.A. Thibault, K.R. Fast, V.T.K. Sauer, J.E. Losby, Z.Diao, M. Belov, and M.R. Freeman
\\
(Dated: 04 Feb 2022)

\beginsupplement

\section{Sample Fabrication}
\label{supp:sect:Fabrication}

Primary considerations for the resonator design are paddle size, mechanical constants of the torsion arms, resonant frequency, torque sensitivity, and depth of undercut that forms a Fabry-Perot optical cavity between the resonator and the substrate base. Two sets of samples were fabricated. Each set has 120 samples that are organized into 6 arrays with 20 samples in each array. 

The magnetic bilayers were patterned into discs having a diameter of 1.36 $\si{\mu}$m by electron beam lithography and lift-off.  The magnetic thin film deposition was done using a confocal magnetron sputtering system (ATC Orion 8, AJA International) with 2-inch diameter targets (Plasmaterials, Inc.)  The chamber base pressure was below 0.2 microTorr. The first layer, in contact with the resonator paddle, is antiferromagnetic cobalt oxide that was reactively sputtered from a cobalt target (deposition rate of 0.228 nm/min using RF power 118 W, sputtering gas pressure 4 mTorr, flow rates of 14 SCCM Ar and 6 SCCM O$_2$). Before depositing the ferromagnetic permalloy layer, the sputtering chamber was purged of oxygen by pressure cycling with argon (filling to argon pressure of 40 mTorr, holding for 60 seconds and then pumping down for 60 seconds, all repeated five times). The argon purge ensures that an antiferromagnetic oxide does not form on the surface of the FM layer (torque effects of NiO on permalloy were discovered by Prosen et al. in the early 1960s\cite{prosen1961}). The permalloy layer was deposited on top of the cobalt oxide layer by DC sputtering from nickel and iron targets (40 W and 13 W powers, respectively, yielding deposition rates of 1.13 nm/min and 0.31 nm/min at 4 mTorr with a 20 SCCM argon gas flow rate). Both layers are 20 nm thick. The densities were assumed to be: 6.44 g/cm$^3$ for the cobalt oxide, 8.91 g/cm$^3$ for the nickel, and 7.86 g/cm$^3$ for the iron. A quartz crystal monitor was used to determine the deposition rates.  

\section{Measurement Apparatus}
\label{supp:sect:Apparatus}

The apparatus was designed to investigate the magnetic properties of nanoscale samples. The primary components of the apparatus are a cryostat, cryogen and vacuum systems, DC magnet and positioning system, Gauss meter, He-Ne laser, steering optics, three-axis objective positioning stage, photoreceiver, High Frequency RF lock-in amplifier (Zurich Instruments HF2LI), RF amplifier, control computer, and power supplies. Coarse adjustment of the cryostat temperature is done by manual control of the helium flow valves. Fine ($\pm$ 0.05 K) temperature stabilization is provided by PID control of a heater that is mounted in the cryostat. The system will stabilize at any desired operating point from helium temperature to 300 K. The apparatus is assembled on a standard optics table and is housed in an enclosure that provides stability of the ambient temperature. 

\subsection{DC Field}

\begin{figure}[htb!] 
	\centering
	\includegraphics[width=0.5\columnwidth]{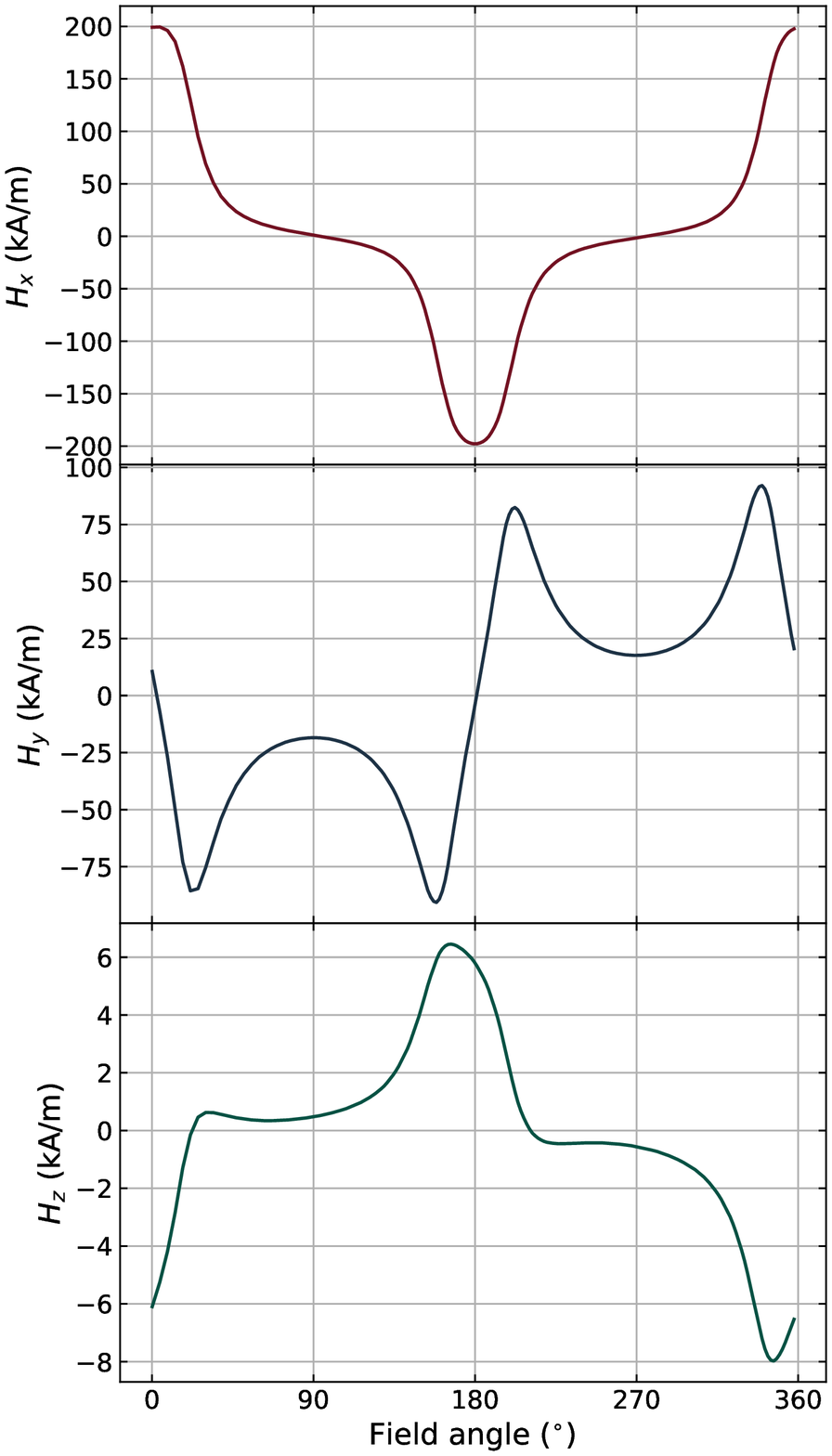}
	\caption{The vector DC magnetic field calibration at the bias magnet position used for the measurements of Figs. 2, 3, and 4, plotted against in-plane field angle.}
	\label{fig:fieldcal}
\end{figure}

A NdFeB permanent magnet (N42 alloy, $6"\times 2"\times 2"$ rectangular prism) is used for applying the DC bias magnetic field. This choice is dictated by the need to avoid a proximal source of variable heat (as would be generated by an electromagnet) in order to maintain temperature stability of the apparatus.  The permanent magnet has sufficiently strong external fields to suit a variety of applications: setting exchange bias, fixed field direction hysteresis measurements (with just a small window of inaccessible fields near zero bias), and rotational measurements where the in-plane field direction rotates through 360 degrees.  The drawback of this approach is that the in-plane field magnitude does not remain constant as the magnet rotates.  The magnet is rectangular thus causing the field strength variation to become larger as the centre of the magnet moves closer to the sample. The maximum and minimum in-plane fields were 199.8 kA/m at 359.6$^{\circ}$ and 17.7 kA/m at 93$^{\circ}$ respectively.   A detailed field calibration was performed using a Hall probe at the sample position to measure each component of magnetic field during magnet rotations and translations such as are executed during the experimental measurements. The results of the calibration for the field rotation measurements are shown in Figure \ref{fig:fieldcal}.  Macrospin simulations of field rotations were conducted using an interpolated version of the field calibration data to ensure that the simulations reflected experimental conditions as closely as possible. 

\subsection{RF Field}

\begin{figure}[htb!] 
	\centering
	\includegraphics[width=1\columnwidth]{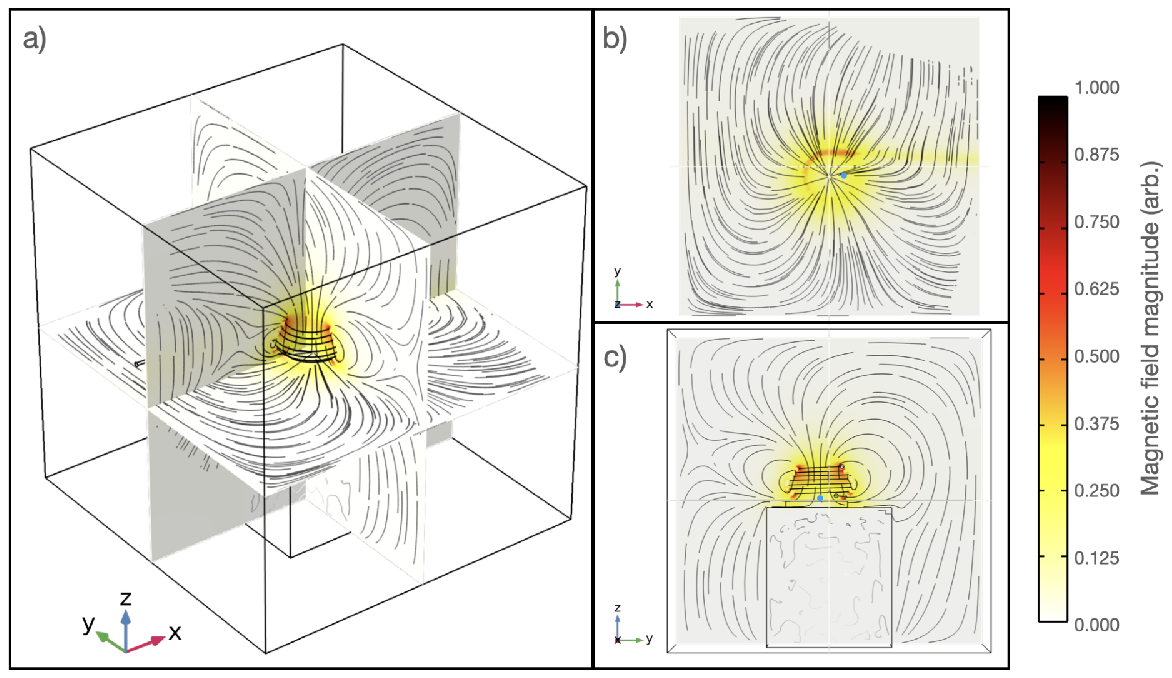}
	\caption{COMSOL\cite{COMSOL} simulations of RF field. There is a significant effect of screening that occurs due to the position of the silicon sample chips placement upon the cryostat cold finger. In panels b) and c), a blue dot is used to signify the approximate position of the sample beneath the coil.}
	\label{fig:ACfieldsim}
\end{figure}

Another matter of concern for three axis measurements of torque is the RF field strength along each Cartesian axis. The driving field in this case is generated by a hand-wound air core RF coil suspended above the sample. The coil winding was not uniform. The devices were offset from the coil axis, so as to produce a nonuniform field for driving multiple torque axes. As Figure \ref{fig:ACfieldsim} indicates, the driving field was strongest along the $z$ and $x$ direction and a very small driving field was expected along $y$. We modeled the coil using AutoDesk\texttrademark  Inventor and COMSOL's\cite{COMSOL} Live Link feature. We simulated the field strength at a location corresponding to the sample position. Images of the sample and coil assembly were used to obtain the coil dimensions, relative position of the sample, and to bracket uncertainty in sample position. The results of the COMSOL study revealed the relative field strengths per Ampere of current. The field strengths were $H_{x}^{\textrm{RF}}=(270 \pm 20)$ (A/m)/A, $H_{y}^{\textrm{RF}}=(0 \pm 20)$ (A/m)/A, and $H_{z}^{\textrm{RF}}=(350 \pm 20)$ (A/m)/A. The results of the simulation are shown in Figure \ref{fig:ACfieldsim}. The blue dots in panels b) and c) represent the relative position of the sample beneath the coil. The simulated driving field strengths were used for calculating the three torque components. 

\subsection{Torque Generation}

The DC bias field and RF dither fields combine in cross-products of the moments and external fields as follows, 
\begin{equation}
    \mathbf{\tau} = \mu_0 \mathbf{m} \times \mathbf{H}
\end{equation}
The individual torque components are,
\begin{align}
    \tau_x &= \mu_0 (m_yH_z - m_zH_y)  \\ 
    \tau_y &= -\mu_0 (m_xH_z - m_zH_x)  \\
    \tau_z &= \mu_0 (m_xH_y - m_yH_x) 
\end{align}
The moments and fields have DC and RF components. Looking at $\tau_y$, we can expand (3) to obtain, 
\begin{align}
    \tau_y = & -\mu_0[(m_x^{DC}+m_x^{RF})(H_z^{DC}+H_z^{RF}) -(m_z^{DC}+m_z^{RF})(H_x^{DC} + H_x^{RF})]
\end{align}
Grouping terms,
\begin{align}
    \tau_y = -\mu_0 &(m_x^{DC}H_z^{DC} + m_x^{DC}H_z^{RF}+ m_x^{RF}H_z^{DC} + m_x^{RF}H_z^{RF} \nonumber \\&- m_z^{DC}H_x^{DC} - m_z^{DC}H_x^{RF} - m_z^{RF}H_x^{DC} - m_z^{RF}H_x^{RF})
\end{align}
The lock-in amplifier only measures the component at the drive frequency.  This leaves
\begin{align}
    \tau_y = -\mu_0 &(m_x^{DC}H_z^{RF} + m_x^{RF}H_z^{DC} - m_z^{DC}H_x^{RF} - m_z^{RF}H_x^{DC} )
\end{align}
At saturation for the $y$-torque, the $z$-moment can be described by the susceptibility, while the $x$-moment is saturated thus substantially reducing its dependence on susceptibility and field. This yields a new form of the $y$-torque,
\begin{align}\label{eq:tauy}
    \tau_y = -&\mu_0[ m_x^{DC}H_z^{RF} + m_x^{RF}H_z^{DC} -\chi_zVH_z^{DC}H_x^{RF} - \chi_zVH_z^{RF}H_x^{DC}]
\end{align}
where the susceptibility $\chi_z$ is given as 
\begin{equation}
    \chi_z = \frac{M_s}{H_{x,\textrm{eff}}+H_{x,\textrm{appl}}}
\end{equation}
with the effective field defined by the demagnetizing factors, $H_{\textrm{eff}} = (N_z -  N_r)M_s$. Equation 8 can be generalized to all torque axes. 

\section{Thermomechanical Calibration}
\label{supp:sect:Calibration}

The thermomechanical calibration enables an absolute scale to be applied to the magnetically-driven mechanical torques and, in the present work, must be performed for three torsion axes. The calibration procedure is the same for each axis.  The average potential energy of a torsion spring is given by,
\begin{equation}
    \langle U\rangle=\frac{1}{2}\kappa^{\textrm{eff}}_{i} \langle \theta_{i}^{2} \rangle,
\end{equation}
where $\kappa^{\textrm{eff}}_{i}$ is the effective torsional spring constant for torsion axis $i$, and $\langle \theta^{2}_{i} \rangle$ is the average angle squared for an induced torque along the axis $i=(x,y,z)$. Utilizing the harmonic relation between angular frequency, moment of inertia and the effective spring constant, we used finite element analysis software\cite{COMSOL} to model our device and perform an integration over the entire sample volume to obtain the moment of inertia for a given mode. 

In the absence of an alternating magnetic field, the torque on the device is due to random thermal fluctuations, the frequency-independence of Brownian motion providing a broadband drive\cite{losby2012}. We use the equipartition of energy to equate the rotational energy and the thermal energy at temperature $T$. From this we find,
\begin{equation}\label{eq:avgtheta}
    \langle \theta^{2}_{i} \rangle=\frac{k_{B}T}{\kappa^{\textrm{eff}}_{i}},
\end{equation}
where $k_{B}$ is Boltzmann's constant. Under the assumption of small angles, we may modify the above equation to give the mean displacement by utilizing the distance from the axis of torque to the location of detection, $\langle x_{i}^{2} \rangle=R_{i}^{2}\langle \theta_{i}^{2} \rangle$. The mechanical response to such a thermal drive is described by a Lorentzian from which the frequency dependent angular spectral density, $S_{\theta_{i}}(f)$, can be obtained in units of rad$^{2}$/Hz for a mechanical device with torsional eigenfrequency $f_{i}$ and quality factor $Q_{i}$,
\begin{equation}
    S_{\theta_{i}\theta_{i}}(f)=\frac{2k_{B}Tf_{i}^{3}}{\pi \kappa_{i}^{\textrm{eff}}Q_{i}} \frac{1}{(f_{i}^{2}-f^{2})^{2}+(\frac{ff_{i}}{Q_{i}})^{2}} .
\end{equation}
We may substitute \ref{eq:avgtheta} into this equation, and by taking the peak spectral density ($S_{\theta_{i}\theta_{i}}$ is maximized for $f = f_{i}$) and multiplying by $R_{i}^{2}$ we find the position spectral density, $S_{x_{i}x_{i}}$,
\begin{equation}
    S_{x_{i}x_{i}}=\frac{2Q_{i}\langle x_{i}^{2}\rangle}{\pi f_{i}}.
\end{equation}
From this equation, the thermomechanical torque spectral density is obtained using $\langle\tau_{i}\rangle=\kappa^{\textrm{eff}}_{i}\langle \theta_{i}\rangle$ and also scaling by $Q_{i}^{-1}$ to account for the enhancement of displacement on mechanical resonance,
\begin{equation}
    S_{\tau_{i}}=\frac{\kappa_{i}^{\textrm{eff}}\sqrt{\langle x_{i}^{2}\rangle}}{R_{i}Q_{i}}.
\end{equation}

Finally, an analysis was performed to find the voltage spectral density from measured data $S_{V_{i}V_{i}}$ (temporal), corresponding to the square of the peak height minus the square of the technical noise floor, divided by the bandwidth of the lock-in measurement,
\begin{equation}
    S_{V_{i}V_{i}}=\frac{V^{2}_{i,\textrm{peak}}-V^{2}_{i,\textrm{background}}}{f_{\textrm{BW}}}.
\end{equation}

The calibration factors, $C_{\tau_{i}}$, were obtained by dividing the torque spectral densities by the square roots of the voltage spectral densities. The resulting factors are in units of Nm/V.  In Table S1, the calibration constants are presented using the more natural scale of aNm/mV for the present measurements. The uncertainty in the thermomechanical calibration, resulting from an inadequate signal to noise ratio of required thermomechanical data near the apparatus noise floor, prevented any significant reduction in the overall uncertainty in the measurements. The second largest source of uncertainty was the RF field magnitude as mentioned in Section S2 B. The combination of these uncertainties results in a $20\%$ systematic uncertainty in calibrated values of torque. 
\begin{table}[h]
    \centering
    \begin{tabular}{| c | c | c |}
    \hline 
       $C_{\tau_{i}}$  & Conversion at 295 K (aNm/mV) & Conversion at 69 K (aNm/mV)\\
       \hline\hline
       $C_{\tau_{x}}$  & 8.6$\pm$0.4 & 2.67$\pm$0.05\\
        $C_{\tau_{y}}$ & 5.0$\pm$0.5  & 2.90$\pm$0.09 \\
        $C_{\tau_{z}}$ & 11.1$\pm$0.6  & 6.4$\pm$0.4 \\
        \hline
    \end{tabular}
    \caption{Torque calibration constants obtained for the sample at ambient and cryogenic temperatures.}
    \label{tab:my_label}
\end{table}

\section{In-plane torque at low bias field}

\begin{figure}[h!]
    \centering
    \includegraphics[width=1\columnwidth]{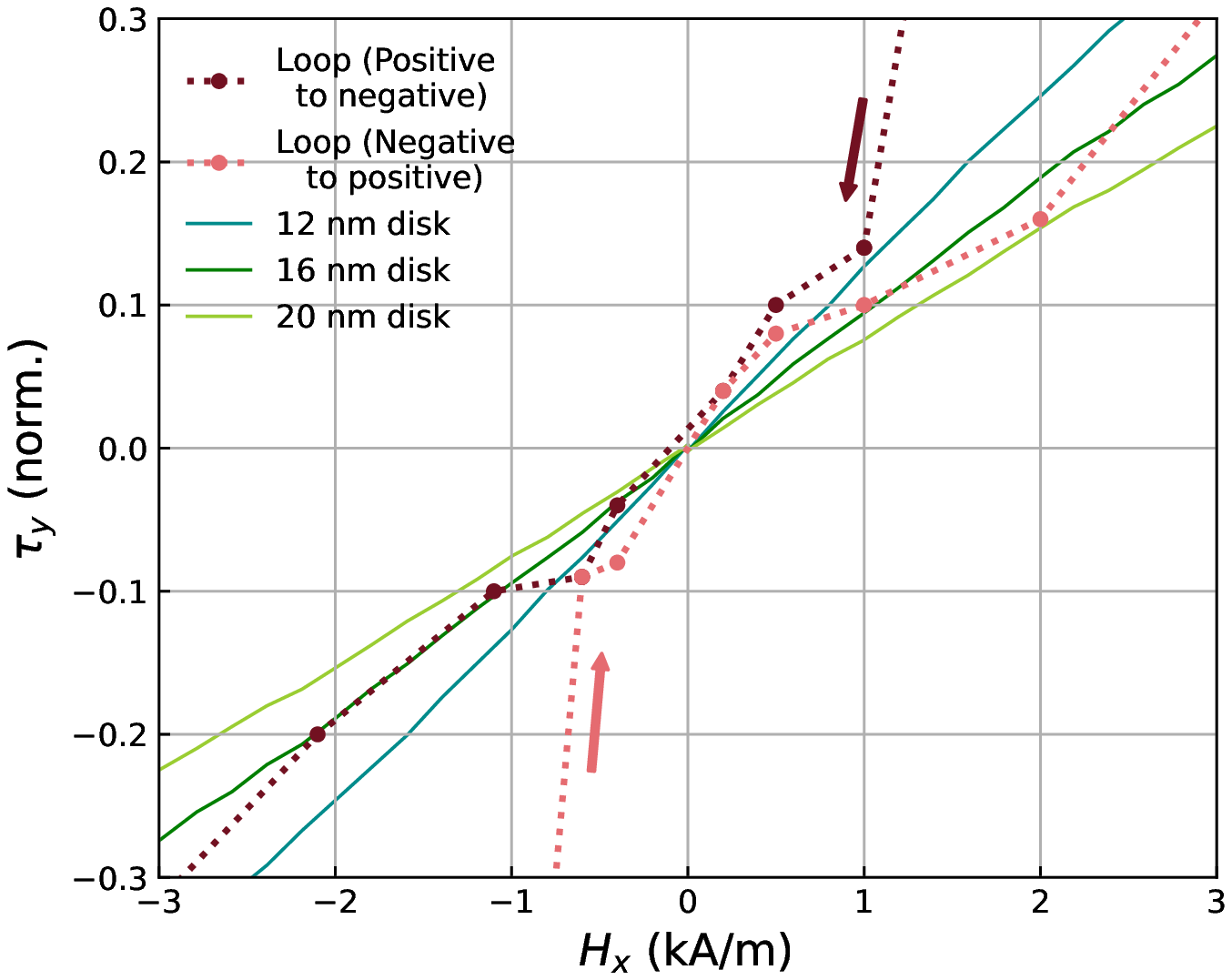}
    \caption{In low external field strengths at room temperature, the permalloy disk demagnetizes, adopting a vortex spin texture. The experimental signature of vortex nucleation is indicated in the above plot by arrows. Mumax3 micromagnetic software \cite{vansteenkiste2014} was used to simulate the vortex behaviour, and the results served to bracket the uncertainty in the magnetically active volume of the disk.}
    \label{fig:thickness_bracket}
\end{figure}

At room temperature in low bias field, and owing to constraints on the size of our magnetic sample, the permalloy disk demagnetized to form a vortex spin texture. Details of the vortex state in this particular magnetic sample at room temperature have been previously discussed \cite{Fast2021}. Experimental data of linear hysteresis loops, as shown in Figure \ref{fig:thickness_bracket}, revealed the field where the vortex spin texture nucleated at sufficiently low bias field strength. Figure \ref{fig:thickness_bracket} shows the lowest field regime of a linear hysteresis loop. The experimental data is represented by solid points that are connected by dotted lines as a guide to the eye. The arrows next to the lines show the field sweep direction, and the section of line they are next to represents the irreversible decrease in $y$ torque that marked the field where vortex nucleation occurred. The in-plane torque in low field followed a linear trend with occasional discrete steps due to the Barkhausen effect, related to the polycrystallinity of the sample \cite{Fast2021}. The slope of this linear trend depends greatly on the ferromagnets aspect ratio since the only anisotropy present at room temperature is due to the shape of the magnetic sample. Three mumax3 micromagnetic simulations \cite{vansteenkiste2014} were performed for comparison with data, and bracketed the uncertainty in the disk thickness. Performing these simulations with mumax3 was essential since it did not require a priori knowledge about the demagnetizing factors. The permalloy disk thickness was determined to be (16$\pm$1) nm. 

We additionally used mumax3 simulations to ensure that the gyrotropic mode of the vortex state was sufficiently above the mechanical resonance, and that we could therefore assume a uniform susceptibility. Results from mumax3 gave a gyrotropic mode frequency of 100 MHz, well above the highest mechanical resonance frequency (4.23 MHz). 

\section{In-plane torque at high bias fields and experimental justification for the macrospin description}

In this section we describe how $M_s$ was obtained from the field-dependence of the in-plane torque in magnetic saturation. 
Non-zero torques on specimens arise when there is an energy cost or gain from a magnetic moment rotating in response to a changing direction of applied field.  If we describe a change in field direction by the addition of a small component, $\delta H_{\perp}$, perpendicular to an existing field, $H_0$, then in the complete absence of anisotropy the magnetic susceptibility, $\chi_{\perp}$, to $\delta H_{\perp}$ will be such that 
\begin{equation}\label{eq:S5torque}
    \mu_0 m_0 \delta H_{\perp} - \mu_0 \chi_{\perp} \delta H_{\perp} H_0 = 0. 
\end{equation}
 In other words, there is no torque perpendicular to the plane defined by $m_0$ (parallel to $H_0$ in the absence of anisotropy) and $\delta H_{\perp}$.  

Qualitatively, the shape of the torque vs. applied field curves in Fig. 1 of the article is readily understood through the cross-products of magnetization and applied field.  Strong shape anisotropy makes the out-of-plane $\chi_{\perp}$ small and nearly constant over this field range (in the thin disk limit, the out-of-plane demagnetization factor, $N_z$, is close to one), whereas the structure magnetizes much more easily in-plane .  The out-of-plane susceptibility decreases significantly only for applied fields on the scale of the demagnetizing field, $O$(400 kA/m) in this case, when the Zeeman energy density becomes comparable.  At low fields, the first term of equation \ref{eq:S5torque} grows rapidly with $H_0$ while the second term remains small.  At higher fields, after $m_0$ saturates, the second term continues to increase in magnitude as $H_0$ rises, causing the net torque to decrease.  $\chi_{\perp}$ depends only on the demagnetization factors and saturation magnetization, $M_s$.  The shape of the specimen is known from the nanofabrication steps and therefore $M_s$ can be deduced from the measured curves.  We extract $M_s$ from comparisons with micromagnetic simulations\cite{vansteenkiste2014} of torque thus avoiding the use of approximations such as the assumption of constant $\chi_{\perp}$.  

In addition, we neglect the possibility of uniaxial perpendicular anisotropy affecting the determination of $M_s$ because our samples are cooled with the fields applied in-plane. PMA in perpendicular field-cooled permalloy/CoO thin films is discussed by Zhou \emph{et al.} \cite{Zhou2004}  Scaling from the maximum positive interface anisotropy of 0.4 mJ/m$^2$ induced by Zhou \emph{et al.} by cooling CoO/Py multilayers in perpendicular fields of 1.6 MA/m, to our samples if hypothetically cooled in strong perpendicular field, still leaves the high-field slope of the in-plane torque curve heavily dominated by the shape anisotropy but would require a 2\% correction to our determination of $M_s$ at liquid nitrogen temperatures.  Micromagnetic simulations show that a positive PMA of 13 kJ/m$^{3}$ would, if neglected, cause a 2\% underestimate of $M_s$. 

In principle, there is redundancy inherent in this determination since the overall torque magnitude is also proportional to $M_s$.  However, for the present experiment the combined uncertainties in thermomechanically-calibrated torque sensitivity and in RF drive field strength at the sample position make the torque magnitude determination of $M_s$ less accurate than the high-field slope analysis, which gave values of (763$\pm$7) kA/m at 295 K and (801$\pm$8) kA/m at 68 K.

\begin{figure}[htb!] 
	\centering
	\includegraphics[width=0.5\columnwidth]{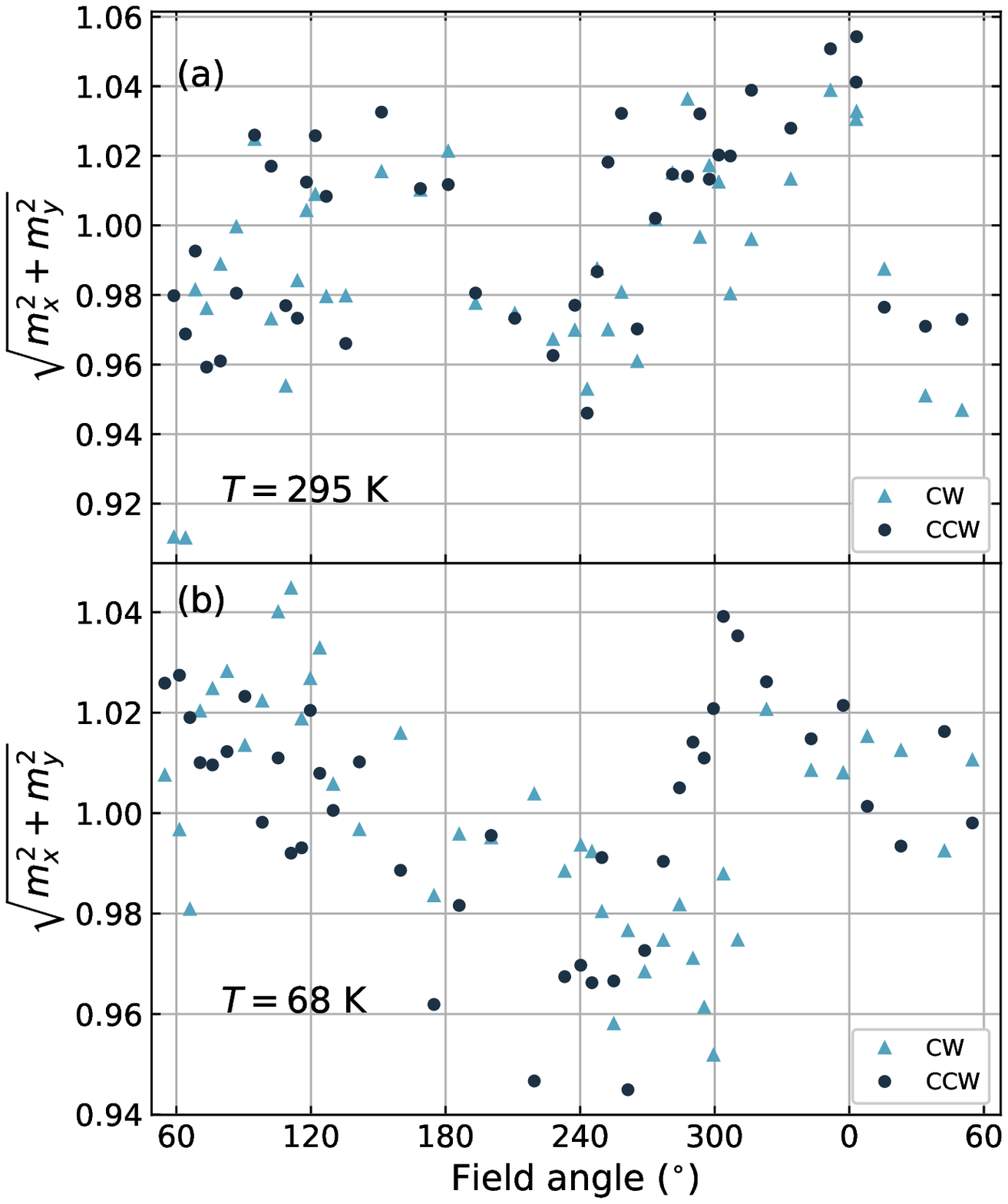}
	\caption{Quadrature addition of magnetic moment components reveals that addition of the in plane magnetic moment components results in near unity for all angles at both ambient (a) and cryogenic (b) temperatures.}
	\label{fig:mxplusmy}
\end{figure}

After the high-field torque slopes are established, it is straightforward to convert the measured $x$- and $y$-torques into $y$- and $x$-magnetic moments, respectively.  The sums of the squares of the in-plane moments then yields a check on the constancy of the magnetic moment during field rotation.  The results for the room temperature and $T=68\thinspace$K data from Fig. 2 are shown in Figure \ref{fig:mxplusmy}.     

\section{LLG macrospin simulations}

\subsection{The LLG equation}

We employ the continuum approximation for cases in which atomistic models cannot predict effects of condensed matter systems. In this regime, the magnetic moments of atoms within a condensed matter system are averaged, and their contributions to magnetization are treated classically. The Landau-Lifshitz-Gilbert equation is a differential equation that describes magnetization dynamics within the continuum approximation. The dynamical equation is as follows,
\begin{equation}
    \frac{d\vec{M}}{dt}=-\gamma \vec{M}\times \vec{B}_{\textrm{eff}}+\frac{\alpha}{M_{S}}\vec{M}\times \frac{d\vec{M}}{dt},
\end{equation}
where $\vec{M}$ is the magnetization, $\gamma$ is the gyromagnetic ratio, $\vec{B}_{\textrm{eff}}$ is the effective field, $\alpha$ is the Gilbert damping term, and $M_{S}$ is the saturation magnetization. The LLG equation contains two terms: a precession term (coefficent $\gamma$) and a damping term (coefficent $\alpha / M_{S}$). If we assume spherical symmetry ($M_{r}=M_{S}$) and define the effective field as the functional derivative of energy density, ($\varepsilon$), by magnetization ($\vec{B}_{\textrm{eff}}=-\frac{\delta \varepsilon}{\delta\vec{M}}$), we can rewrite the equation as a coupled system of differential equations in terms of polar ($\phi$) and azimuthal ($\theta$) angles,
\begin{equation}
   \begin{aligned}
     \frac{d\theta}{dt}&=-\frac{\gamma}{\mu_{0}M_{s}(1+\alpha^{2})}\Big(\frac{1}{\sin (\theta )}\frac{\partial \varepsilon}{\partial \phi}+\alpha\frac{\partial\varepsilon}{\partial \theta}\Big)
     \\
     \frac{d\phi}{dt}&=\frac{\gamma}{\mu_{0}M_{s}(1+\alpha^{2})}\Big( \frac{-\alpha}{\sin ^{2} (\theta )}\frac{\partial \varepsilon }{\partial \phi}+\frac{1}{\sin (\theta )}\frac{\partial \varepsilon}{\partial \theta} \Big).
    \end{aligned}
\end{equation}
Under conditions of sufficiently long time scales, the damping term will dominate the precession term and the magnetization will approach equilibrium. Our experimental frequencies were in the range of several MHz, not fast enough to meet the timescales of even the slowest dynamics predicted by the LLG equation. As such, we obtained a solution to the LLG equation and recorded the magnetization direction as polar and azimuthal angles in the portion of the solution dominated by damping. To calculate the AC torque, a dither field was applied to the field direction. Most of the AC field was along the $z$ and $x$ directions, while the AC field along $y$ was very small in comparison owing to the position of the single RF excitation coil. COMSOL\cite{COMSOL} simulations of field strength at the sample location were further corroborated by expected locations of zero crossings in the rotating data from our room temperature study of permalloy. The externally applied AC magnetic field is increased in the $z$ and $x$ directions by a small amount to a maxima, then decreased below the initial field to a minima, and finally increased back to it's original applied field strength. Following the dither steps, the DC field is then stepped to the next field angle where the process begins again. For each dither point, the torque was calculated in the usual way, $\vec{\tau}=\mu_{0}\vec{m}\times\vec{H}_{\textrm{ext}}^{\textrm{DC}}$. A linear fit was then applied to the torque response. The AC torque can be deduced from the slope of the fit.

\subsection{Rotatable anisotropy}

For most anisotropies, calculation of torque is straightforward. All that is required is to specify the strength of the anisotropy (by way of some anisotropy constant) and the angle at which the anisotropy axis is directed. For the rotatable anisotropy, however, extra care must be taken. The existence of the rotatable anisotropy axis is due to the exchange coupled uncompensated spins at the FM/AFM interface\cite{mcmichael1998,stiles1999}. The energy density is written in the following way,
\begin{equation}
    \varepsilon_{\textrm{rot}}=- K_{\textrm{rot}} \hat{M}_{\textrm{FM}} \cdot \hat{M}_{\textrm{rot}},
    \label{eqn:rot}
\end{equation}
where $K_{\textrm{rot}}$ is the rotatable anisotropy constant, $\hat{M}_{\textrm{FM}}$ is the unit vector of the FM magnetization, and $\hat{M}_{\textrm{rot}}$ is the average direction of the uncompensated spins. Since these spins are able to rotate, they will follow the magnetization as it rotates with the applied field direction, however they may deviate slightly. To account for deviation we introduced a relative angle $\Delta\Phi_{\textrm{rot}}$ that defines the relative angle between the FM magnetization direction and the interfacial spins. We can rewrite equation \ref{eqn:rot} as,
\begin{equation}
    \varepsilon_{\textrm{rot}}=- K_{\textrm{rot}} \cos (\Delta\Phi_{\textrm{rot}}),
    \label{eqn:rot2}
\end{equation}
which produces an easy axis along the direction $\Delta\Phi_{\textrm{rot}}$. If we apply a small perturbation ($\Delta\alpha$) to the angle of the FM magnetization, we find,
\begin{equation}
    \varepsilon_{\textrm{rot}}=- K_{\textrm{rot}} \cos (\Delta\Phi_{\textrm{rot}}-\Delta\alpha).
    \label{eqn:rot3}
\end{equation}
There is a nonzero curvature in the energy density that will contribute to the AC torque. As such, when the DC field is stepped, there is no contribution from the rotatable anisotropy, but on subsequent AC dither cycles, the rotatable anisotropy term is included and contributes to the torque. 

\subsection{Additional contributing anisotropies}

Exchange coupling between FM and AF layers introduces rotatable, spin-flop and unidirectional anisotropy terms given by the following energy density equation
\begin{equation}
   \begin{aligned}
     \varepsilon_{\textrm{A}}=&-K_{\textrm{SF}}\sin^{2}(\theta)\cos^{2}(\phi-\phi_{\textrm{SF}})-K_{\textrm{EB}}\sin(\theta)\cos(\phi-\phi_{\textrm{EB}})-K_{\textrm{rot}}\hat{m}_{\textrm{FM}}\cdot\hat{H}_{\textrm{rot}}.
    \end{aligned}
\end{equation} 
Aside from the rotatable anisotropy (described in subsection S6 B), implementation of spin-flop and unidirectional anisotropies is straightforward. 

The shape anisotropy is obtained from equations derived by Joseph\cite{joseph1966} to calculate the cylindrical demagnetization factors. The demagnetizing energy density is thus, 
\begin{equation}
    \begin{aligned}
     \varepsilon_{\textrm{D}}=\frac{1}{2}\mu_{0}M_{S}^{2}\big(& N_{x}\cos^{2}(\phi)\sin^{2}(\theta)+N_{y}\sin^{2}(\phi)\sin^{2}(\theta)+N_{z}\cos^{2}(\theta) \big)
    \end{aligned}
\end{equation}  
where $N_{i}$ is the demagnetizing factor along axis $i$. Finally we introduce the Zeeman energy density that describes the effect of an external field on the macrospin
\begin{equation}
   \begin{aligned}
     \varepsilon_{\textrm{Z}}=&-\mu_{0}\vec{m}_{\textrm{FM}}\cdot \vec{H}_{\textrm{ext}}^{\textrm{DC}}.
    \end{aligned}
\end{equation} 

\subsection{Curve fitting}

A unique match of the data to the multi-parameter model is possible because each anisotropy parameter affects specific features of the AC torque curve independent of the other anisotropies. We presume that the shape anisotropy is identical at both room temperature, and cryogenic temperatures. The demagnetizing energy density has a significant effect on the $\tau_{x}$ and $\tau_{y}$ curves, however since we are considering a cylindrical sample $N_{x}=N_{y}$, no $\tau_{z}$ signal is observed as a result of shape anisotropy. As a consequence, all components of the $\tau_{z}$ curve are derived from anisotropies that emerge due to exchange coupling between the FM and AF layers. We introduce three anisotropies that are a direct result of exchange coupling: the rotatable anisotropy that represents the effect due to exchange coupled compensated AF spins, unidirectional anisotropy that represents exchange bias, and uniaxial anisotropy that represents spin-flop. The rotatable anisotropy term produces the 1-fold rotational symmetry that is dominant in the low temperature $\tau_{z}$ data. The rotatable anisotropy magnitude accounts for the slope of the linear regions around the zero crossing. The unidirectional exchange bias anisotropy term, at the field cooling angle presented in Figure 3 of the manuscript, contributes an extrema height asymmetry. The angle that the unidirectional anisotropy is applied along can specifically suppress the height of one of the two extrema that come from the contribution of the rotatable anisotropy. Finally, the uniaxial anisotropy accounts for the contribution of canting spins in the AF layer (spin-flop), which reduces the magnitude of the peaks due to the rotatable anisotropy. 

Fitting the low temperature $z$-torque data to the model yielded a minimum $\chi^2_{\nu,\textrm{min}}$ value of 1.2 where $\chi^{2}_{\nu}$ was calculated from
\begin{equation}
    \chi^{2}_{\nu}=\frac{1}{\nu}\sum_{i}\frac{(\tau^{\textrm{exp}}_{i}-\tau^{\textrm{model}}_{i})^{2}}{(\delta\tau_{i}^{\textrm{exp}})^{2}},
\end{equation}
where $\tau^{\textrm{exp}}_{i}$ is the value of the i-th data point, $\tau^{\textrm{model}}_{i}$ is the value of the i-th point of the macrospin solution, $\nu$ is the degrees of freedom (number of data points minus the number of fit parameters), and $\delta\tau_{i}^{\textrm{exp}}$ is the uncertainty in the i-th data point. Each parameter is varied around its best-fit value to obtain $\chi^{2}_{\nu}+\delta\chi^{2}_{\nu}$ where $\delta\chi_{\nu}^{2}=\sqrt{2\nu}$. In the present work $\chi^{2}_{\nu,\textrm{min}}=1.2$ and $\delta\chi^{2}_{\nu}=0.3$ so each parameter ($K_{\textrm{EB}}$, $\phi_{\textrm{EB}}$, $K_{\textrm{SF}}$, $\phi_{\textrm{SF}}$, $K_{\textrm{rot}}$, $\Delta\Phi_{\textrm{rot}}$) is varied such that $\chi^{2}=1.5$, thus bracketing the 1 sigma confidence interval of each parameter individually. We find $K_{\textrm{EB}}=(0.10\pm 0.05)$ kJ/m$^{3}$, $\phi_{\textrm{EB}}=55^{\circ}\pm 3^{\circ}$, $K_{\textrm{SF}}=(8.3\pm 0.3)$ kJ/m$^{3}$, $\phi_{\textrm{SF}}=-91^{\circ}\pm 2^{\circ}$, $K_{\textrm{rot}}=(22 \pm 1)$ kJ/m$^{3}$, and $\Delta \Phi_{\textrm{rot}}=-5^{\circ}\pm3^{\circ}$.

\subsection{Dynamics}

Determination of magnetization dynamics is not the only way to study anisotropic systems, especially for long timescales without dynamics. An analysis was also performed by using the Stoner-Wohlfarth model in 2D to obtain the out of plane torque. Good agreement between Stoner-Wohlfarth and LLG simulation methodologies was observed. While 3 dimensional extensions of the Stoner-Wohlfarth model can also be used to model 3 axis torques, the LLG equation was used for analysis of high field hysteresis loops to bracket $M_{S}$ and the magnetic aspect ratio. Further studies of micromagnetic exchange bias will undoubtedly benefit from LLG micromagnetic simulation\cite{de2016modelling,de2017modelling,declercq2017thesis}. Presenting a LLG macrospin model that agrees with models such as Stoner-Wohlfarth serves as an appropriate jumping off point for subsequent studies of micromagnetic exchange bias wherein micromagnetic domain structure plays an essential part. 
\\
\\
\textit{References for citations in the Supplementary Material section are included in the list on pages 12-14.}
\end{document}